\documentclass[conference]{IEEEtran}
\IEEEoverridecommandlockouts
\usepackage[cmex10]{amsmath}
\usepackage{tikz,enumerate,pgfplots}

\DeclareMathOperator{\perm}{perm}
\newtheorem{lemma}{Lemma}
\newcommand{\qed}{\hspace*{\fill} \fbox{} \par \vspace{\baselineskip}}

\begin{document}
\title{An investigation of SUDOKU-inspired non-linear codes with local constraints}
\author{\IEEEauthorblockN{Jossy Sayir and Joned Sarwar}
\IEEEauthorblockA{University of Cambridge, U.K.
\thanks{Funded in part by the European Research Council under ERC grant
agreement 259663 and by the FP7 Network of Excellence NEWCOM\#
under grant agreement 318306.}
}
}

\maketitle

\begin{abstract}
Codes with local permutation constraints are described.
Belief propagation decoding is shown to require the
computation of permanents, and trellis-based methods
for computing the permanents are introduced.
New insights into the asymptotic performance of such codes
are presented. A universal encoder for codes with local constraints
is introduced, and simulation results for two code structures,
SUDOKU and semi-pandiagonal Latin squares, are presented.
\end{abstract}

\section{Introduction}
The codes we investigate are sets of codewords that
satisfy a number of local non-linear constraints. As such,
they are described by factor graphs just like low-density
parity-check (LDPC) codes. Unlike LDPC codes for which the
constraint nodes enforce linear constraints over fields,
in the codes treated here the constraint nodes enforce
non-linear constraints of the SUDOKU type. Each constraint
requires that {\em all variables involved take on different values
over the code alphabet.} For example, for a code alphabet
$\{0,1,2,3\}$ and a constraint node of degree 4, the variables
involved could take on the values $(2,1,0,3)$ or $(3,0,2,1)$
but not $(2,2,0,1)$ because 2 is repeated. Note that it is
not imperative that the constraint node degree $d_c$ equals the 
alphabet size $q$, but this will be the case in all the structures
we will investigate, in which case the constraint can equivalently
be described as enforcing that the variables involved take
on values over the set of permutations of the code alphabet.

These codes are inspired by SUDOKU puzzles and their study was
initially motivated as a tool for teaching belief propagation
decoding for LDPC codes via its analogy with solving SUDOKU puzzles.
Note that in general, there is no particular reason to arrange the variables
in our codewords in a square $q\times q$ grid as they would be in a
classic SUDOKU puzzle. The reason for arranging them in this
manner is to visualise the constraints corresponding
to rows and columns of the square. In this paper, we will look
at some regular structures that can be visualised as squares
with row-column constraints, but also cover some general structures
that are best represented as a common one-dimenstional array of
variables constrained by  constraint nodes in a factor graph with
random connections. 

In Section~\ref{sec:perm}, we will discuss iterative decoding for 
codes with permutation constraints and show that the operation
in the constraint node is equivalent to computing a set of permanents \cite{cauchy1815}.
We will then show how to compute these permanents using a trellis-based
approach, and specialise this approach for decoding over erasure channels.
In Section~\ref{sec:de}, we will discuss asymptotic performance analysis
for codes with permutation constraints. In Section~\ref{sec:enc}, we 
will present a universal approach to encoding codes with local constraints
and discuss its limitations. In Section~\ref{sec:sim}, we will present
simulation measurements of the performance of two specific code structures,
SUDOKU and semi-pandiagonal Latin squares, over the erasure channel.

\section{Iterative decoding, permanents and trellises}
\label{sec:perm}

\subsection{General belief propagation decoding}

Iterative decoding over factor graphs with non-linear constraints
follows the same rules as iterative decoding for graphs with
linear constraints. The belief propagation algorithm operates
Bayesian estimation under the assumption that messages from the
graph are independent observations, as described in \cite{atkins2014}
and references therein. We will express messages as $q$-ary
probability mass functions, although it may be sensible in calculations
to transfer them to the logarithmic or log-likelihood ratio domain.
For a variable node, the operation is the
same as that of variable nodes for linear codes. For a constraint
node, the operation is 
\begin{equation}
b_{ij} = \xi_i\sum_{(j_1,\ldots,j_i=j,\ldots,j_q) \in S_q} \;\;\prod_{k\neq i} a_{kj_k}.
\label{eq:constraint-op}
\end{equation}
where $a_{ij}$ is the $j$-th component of the $i$-th incoming message to the
constraint node, $b_{ij}$ is the $j$-th component of the $i$-th outgoing
message of the constaint node, $\xi_i$ is a normalisation constant, and $S_q$ is the
symmetric group on $\{1,2,\ldots,q\}$. This can 
also be written as
\begin{equation}
b_{ij} = \frac{\perm(A_{ij})}{\perm(A)},
\label{eq:constraint-op-perm}
\end{equation}
where $A$ is the $q\times q$ matrix of incoming messages. For any
matrix $M$, we write $\perm(M)$ for the permanent of $M$, $m_{ij}$
for the $i,j$-th element of $M$, and $M_{ij}$ for
the matrix obtained by removing the $i$-th
row and $j$-th column from $M$.

Computing a permanent is a high complexity operation. A direct
evaluation of 
\[
\perm(M) = \sum_{(i_1,\ldots,i_q)\in S_q}m_{1i_q}m_{2i_2}\ldots m_{qi_q}
\]
requires $(q-1)q!$ multiplications and $q!-1$ additions.
The best known efficient algorithm for computing an approximation of the permament
of a matrix with positive entries has a probabilistic polynomial
complexity, which polynomial is of degree 11 and does not provide
any benefits for the alphabet sizes of interest to us, i.e.,
$q=9$ or less, perhaps $q=16$. Still, even for $q=9$, the 
number of multiplications is about $3\times 10^6$, which would
need to be computed $q(q^2+1) = 801$ times at each constraint node
at each iteration in order to evaluate (\ref{eq:constraint-op-perm}),
resulting in a total number of multiplication, say for 10 iterations
for a SUDOKU codeword (81 variables, 27 constraints),
of about $6.3\times 10^{11}$ per decoding operation. An efficiently
programmed simulation of a transmission path for a sufficient number
of codewords to measure error performance would hence be beyond the
ability of a modern Gflop computer. 

A solution to this is to compute the permanent using a trellis.
This can be seen as a generalization of Laplace's co-factor expansion
for computing the permanent or determinant. In Laplace's co-factor expansion,
depending on where we start, we  only compute the co-factors in the
first row and the permanent of the matrix, whereas to evaluate (\ref{eq:constraint-op-perm}),
we need all the co-factors, corresponding to repeating Laplace's expansion
starting with every row in turn. 
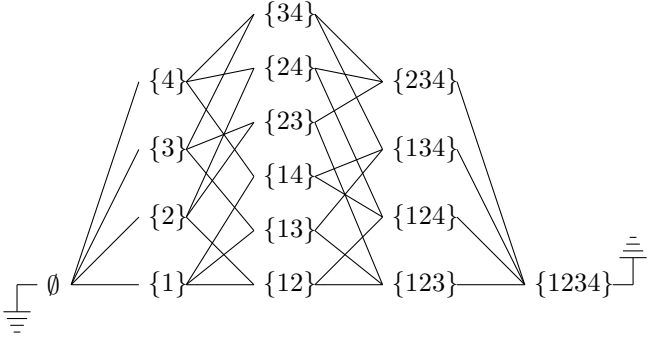
\begin{figure}
\centering
\begin{tikzpicture}[scale=.9]
\draw (0.5,.6) -- (.9,.6);
\draw (.55,.5) -- (.85,.5);
\draw(.6,.4) -- (.8,.4);
\draw(.65,.3) -- (.75,.3);
\draw(.7,.6) -- (.7,1);
\draw(.7,1) -- (1,1);
\node [right] at (1,1) {$\emptyset$};
\draw (1.5,1) -- (2.5,1);
\node [right] at (2.5,1) {$\{1\}$};
\draw (3.2,1) -- (4.2,1);
\node [right] at (4.2,1) {$\{12$\}};
\draw (5.1,1) -- (6.1,1);
\node [right] at (6.1,1) {$\{123\}$};
\draw (7.2,1) -- (8.2,1);
\node [right] at (8.2,1) {$\{1234\}$};
\draw (1.5,1) -- (2.5,2);
\draw (1.5,1) -- (2.5,3);
\draw (1.5,1) -- (2.5,4);
\node [right] at (2.5,2) {$\{2\}$};
\node [right] at (2.5,3) {$\{3\}$};
\node [right] at (2.5,4) {$\{4\}$};
\draw (3.2,1) -- (4.2,1.8);
\draw (3.2,1) -- (4.2,2.6);
\draw (3.2,2) -- (4.2,1);
\draw (3.2,2) -- (4.2,3.4);
\draw (3.2,2) -- (4.2,4.2);
\draw (3.2,3) -- (4.2,1.8);
\draw (3.2,3) -- (4.2,3.4);
\draw (3.2,3) -- (4.2,5);
\draw (3.2,4) -- (4.2,2.6);
\draw (3.2,4) -- (4.2,4.2);
\draw (3.2,4) -- (4.2,5);
\node [right] at (4.2,1.8) {$\{13\}$};
\node [right] at (4.2,2.6) {$\{14\}$};
\node [right] at (4.2,3.4) {$\{23\}$};
\node [right] at (4.2,4.2) {$\{24\}$};
\node [right] at (4.2,5) {$\{34\}$};
\node [right] at (6.1,2) {$\{124\}$};
\node [right] at (6.1,3) {$\{134\}$};
\node [right] at (6.1,4) {$\{234\}$};
\draw (5.1,1) -- (6.1,2);
\draw (5.1,1.8) -- (6.1,1);
\draw (5.1,1.8) -- (6.1,3);
\draw (5.1,2.6) -- (6.1,2);
\draw (5.1,2.6) -- (6.1,3);
\draw (5.1,3.4) -- (6.1,1);
\draw (5.1,3.4) -- (6.1,4);
\draw (5.1,4.2) -- (6.1,2);
\draw (5.1,4.2) -- (6.1,4);
\draw (5.1,5) -- (6.1,3);
\draw (5.1,5) -- (6.1,4);
\draw (7.2,2) -- (8.2,1);
\draw (7.2,3) -- (8.2,1);
\draw (7.2,4) -- (8.2,1);
\draw (9.5,1) -- (9.8,1);
\draw (9.8,1) -- (9.8,1.4);
\draw (9.6,1.4) -- (10,1.4);
\draw (9.65,1.5) -- (9.95,1.5);
\draw (9.7,1.6) -- (9.9,1.6);
\draw (9.75,1.7) -- (9.85,1.7);
\end{tikzpicture}
\caption{Trellis-based permanent computation}
\label{fig:trellis-perm}
\end{figure}
Figure~\ref{fig:trellis-perm} represents the model for the trellis-based
computation of the permanent and all co-factors of a $4\times 4$ matrix.
We start with an empty set at the root of the trellis. Every trellis
stage corresponds to a row of the matrix. At row 1, we can pick any 
of the four elements, resulting in four paths from the empty set to 
the atomic sets $\{1\},\{2\},\{3\}$ and $\{4\}$. At every further
stage in the trellis, you can only advance using columns that have
not been visited yet, so for example there a three paths leading forward
from $\{1\}$ to $\{1,2\}, \{1,3\}$ and $\{1,4\}$. The termination (or ``toor'')
of the trellis corresponds to the set $\{1,2,3,4\}$ where all the columns
have been visited.

Multiplying matrix elements on every edge and adding them in the nodes of
the trellis will compute the permanent of the matrix and the co-factors of its
last row, and is completely equivalent to Laplace's co-factor expansion.
This is also equivalent  to the forward iteration of the BCJR \cite{bcjr} or
forward-backward algorithm. Applying the full BCJR algorithm to the trellis
yields the permanent and all the co-factors we need, which result as the 
sum of the products of the forward sums and the backward sums of all edges corresponding
to an element in a row. For example to compute the co-factor $M_{23}$ using
the trellis in Figure~\ref{fig:trellis-perm}, we need to multiply the forward-sums
and the backward-sums in all the transitions corresponding to a 3, i.e., $\{1\}$
to $\{1,3\}$, $\{2\}$ to $\{2,3\}$, and $\{4\}$ to $\{3,4\}$, then sum those to
obtain $\perm(M_{23})$.

The number of multiplications in the trellis-based permanent computation
excluding the last stage for computing the co-factors is
\[
2\sum_{i=1}^{q-1} {q \choose i} (q-i) = q(2^q-2),
\] 
which is still exponential in the alphabet size, but provides
a significant reduction in complexity at the alphabet sizes of interest
to us. For example, for $q=9$, we now
need to evaluate only 4590 multiplications per constraint node
per iteration, or, repeating the evaluation above for decoding a SUDOKU
codeword, $1.2\times 10^6$ multiplications per decoding operation.

\subsection{Decoding for the erasure channel}

The trellis-based approach described can be further simplified when decoding
for an erasure channel. When decoding for a $q$-ary erasure channel,
messages start up as atomic distributions for non-erased positions,
and uniform $q$-ary distributions for erased positions. Through the
iterations, messages of varying supports will appear but all messages
are always uniform distributions over their support. Hence, we can replace
distribution-valued messages by 0/1 indicators or subsets of possible values.

The subset-valued rule for constraint nodes of an erasure decoder
has been stated in \cite{atkins2014} and we repeat it here. 
Letting $m_{v\to c}(i) \subseteq \{1,\ldots,q\}$ be the incoming subset message to a constraint
node on its $i$-th edge, the resulting constraint node rule for generating the $j$-th
outgoing message $m_{c\to v}(j) \subseteq \{1,\ldots,q\}$ is
\[
m_{c\to v}(j) = \{1,\ldots,q\} - \bigcup_n A_n
\]
where $A_n$ is any set such that
\[
\exists \mathcal{J}\subset \{1,\ldots,q\} \mbox{ such that }
\begin{cases} j\notin\mathcal{J}, \\
A_n=\bigcup_{j'\in\mathcal{J}} m_{v\to c}(j'), \\
\mbox{ and } \#\mathcal{J} =  \#A_n,
\end{cases}
\]
where $\#S$ denotes the cardinality of the set $S$. In
other words, the rule is that whenever the union of a number of
incoming messages has cardinality that number, you can eliminate 
the members of that union from all remaining outgoing messages. 
For example, if incoming messages $i$ and $j$ are both $\{1,2\}$,
then the values 1 and 2 can be reserved for the variables $i$ and $j$
and can be eliminated from all remaining variables in the constraint.
This rule is familiar to those who like solving SUDOKU puzzles.

A direct evaluation of the rule above requires to examine the union of every
combination of incoming messages, resulting in a loop over $2^q$
instances. Again, much can be gained by realizing the constraint
node operation on a trellis. Let $A$ be the matrix of incoming
messages, where $a_{ij}=1$ if element $j$ belongs to the subset
of incoming message $i$ and $a_{ij}=0$ otherwise. Using the same trellis
structure as we did for the evaluation of the permanent, the
following steps are applied:
\begin{enumerate}[1)]
\item Starting from the root node on a trellis with all edges blanked out,
draw a forward path for every edge coming out of a node that satisfies
the conditions
\begin{enumerate}[(i)]
\item the edge exists in the full trellis; and
\item there is a 1 in the matrix corresponding to the trellis stage and
the symbol added by this edge.
\end{enumerate}
\item Prune any paths that did not terminate in the ``toor'' node.
\item Insert a 1 in the outgoing message wherever there is an edge
corresponding to that symbol in the trellis stage for that row.
\end{enumerate}
The method is illustrated for the incoming message matrix
\[
A = \left[
\begin{array}{cccc}
1 & 1 & 1 & 1 \\
1 & 0 & 1 & 0 \\
1 & 1 & 0 & 0 \\
1 & 1 & 0 & 0 
\end{array}
\right]
\]
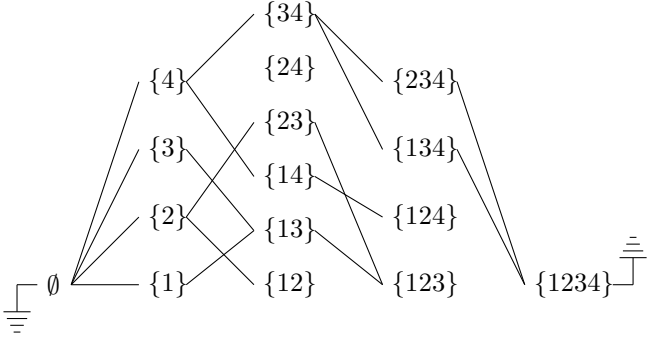
\begin{figure}
\centering
\begin{tikzpicture}[scale=0.9]
\draw (0.5,.6) -- (.9,.6);
\draw (.55,.5) -- (.85,.5);
\draw(.6,.4) -- (.8,.4);
\draw(.65,.3) -- (.75,.3);
\draw(.7,.6) -- (.7,1);
\draw(.7,1) -- (1,1);
\node [right] at (1,1) {$\emptyset$};
\draw (1.5,1) -- (2.5,1);
\draw (1.5,1) -- (2.5,2);
\draw (1.5,1) -- (2.5,3);
\draw (1.5,1) -- (2.5,4);
\node [right] at (2.5,1) {$\{1\}$};
\node [right] at (2.5,2) {$\{2\}$};
\node [right] at (2.5,3) {$\{3\}$};
\node [right] at (2.5,4) {$\{4\}$};
%
\draw (3.2,1) -- (4.2,1.8);
\draw (3.2,2) -- (4.2,1);
\draw (3.2,2) -- (4.2,3.4);
\draw (3.2,3) -- (4.2,1.8);
\draw (3.2,4) -- (4.2,2.6);
\draw (3.2,4) -- (4.2,5);
\node [right] at (4.2,1) {$\{12$\}};
\node [right] at (4.2,1.8) {$\{13\}$};
\node [right] at (4.2,2.6) {$\{14\}$};
\node [right] at (4.2,3.4) {$\{23\}$};
\node [right] at (4.2,4.2) {$\{24\}$};
\node [right] at (4.2,5) {$\{34\}$};
\node [right] at (6.1,1) {$\{123\}$};
\node [right] at (6.1,2) {$\{124\}$};
\node [right] at (6.1,3) {$\{134\}$};
\node [right] at (6.1,4) {$\{234\}$};
%
\draw (5.1,1.8) -- (6.1,1);
\draw (5.1,2.6) -- (6.1,2);
\draw (5.1,3.4) -- (6.1,1);
\draw (5.1,5) -- (6.1,3);
\draw (5.1,5) -- (6.1,4);
%
\draw (7.2,3) -- (8.2,1);
\draw (7.2,4) -- (8.2,1);
\node [right] at (8.2,1) {$\{1234\}$};
\draw (9.5,1) -- (9.8,1);
\draw (9.8,1) -- (9.8,1.4);
\draw (9.6,1.4) -- (10,1.4);
\draw (9.65,1.5) -- (9.95,1.5);
\draw (9.7,1.6) -- (9.9,1.6);
\draw (9.75,1.7) -- (9.85,1.7);
\end{tikzpicture}
\caption{Forward drawing step of the trellis based constraint node operation}
\label{fig:etrellis1}
\end{figure}
\begin{figure}
\centering
\begin{tikzpicture}[scale=0.9]
\draw (0.5,.6) -- (.9,.6);
\draw (.55,.5) -- (.85,.5);
\draw(.6,.4) -- (.8,.4);
\draw(.65,.3) -- (.75,.3);
\draw(.7,.6) -- (.7,1);
\draw(.7,1) -- (1,1);
\node [right] at (1,1) {$\emptyset$};
\draw (1.5,1) -- (2.5,4);
\node [right] at (2.5,1) {$\{1\}$};
\node [right] at (2.5,2) {$\{2\}$};
\node [right] at (2.5,3) {$\{3\}$};
\node [right] at (2.5,4) {$\{4\}$};
\draw (3.2,4) -- (4.2,5);
\node [right] at (4.2,1) {$\{12$\}};
\node [right] at (4.2,1.8) {$\{13\}$};
\node [right] at (4.2,2.6) {$\{14\}$};
\node [right] at (4.2,3.4) {$\{23\}$};
\node [right] at (4.2,4.2) {$\{24\}$};
\node [right] at (4.2,5) {$\{34\}$};
\node [right] at (6.1,1) {$\{123\}$};
\node [right] at (6.1,2) {$\{124\}$};
\node [right] at (6.1,3) {$\{134\}$};
\node [right] at (6.1,4) {$\{234\}$};
\draw (5.1,5) -- (6.1,3);
\draw (5.1,5) -- (6.1,4);
\draw (7.2,3) -- (8.2,1);
\draw (7.2,4) -- (8.2,1);
\node [right] at (8.2,1) {$\{1234\}$};
\draw (9.5,1) -- (9.8,1);
\draw (9.8,1) -- (9.8,1.4);
\draw (9.6,1.4) -- (10,1.4);
\draw (9.65,1.5) -- (9.95,1.5);
\draw (9.7,1.6) -- (9.9,1.6);
\draw (9.75,1.7) -- (9.85,1.7);
\end{tikzpicture}
\caption{Pruning step of the trellis based constraint node operation}
\label{fig:etrellis2}
\end{figure}
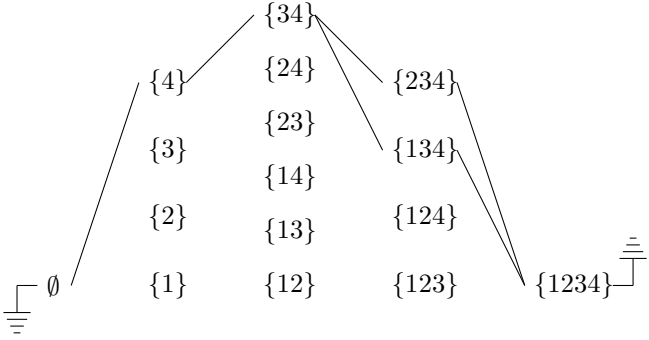
in Figure~\ref{fig:etrellis1} (forward drawing step) and Figure~\ref{fig:etrellis2}
(pruning step), resulting in the outgoing message matrix
\[
B = \left[
\begin{array}{cccc}
0 & 0 & 0 & 1 \\
0 & 0 & 1 & 0 \\
1 & 1 & 0 & 0 \\
1 & 1 & 0 & 0 
\end{array}
\right]
\]
An analysis of the complexity of this method would be difficult as the 
choice of elementary operations to be counted was unclear to us.
Furthermore, the number of operations is not a direct function of the number
of ones in a row as it would vary according to the position of the ones.
In our simulations, we observed an acceleration of several orders
of magnitude by switching from the direct approach of iterating over
the $2^q$ configurations of incoming subsets to the trellis-based
approach, and this acceleration was more pronounced as the alphabet
size increased. On the other hand, the trellis-based approach is 
not easy to implement and requires several pages of code, while
the direct approach is trivial and only a few lines long in most
programming languages.

\section{Asymptotic performance analysis and density evolution}
\label{sec:de}

Density evolution can be applied to analyse the asymptotic performance
of codes with non-linear constraints in a manner similar to LDPC 
codes. The elements of
this analysis have been presented in \cite{atkins2014} and we will not
dwelve on them here. However, we take this opportunity to correct 
two inaccuracies in \cite{atkins2014} and report on new insights obtained
from \cite{vontobel2014} that cast a new light on the results in \cite{atkins2014}.

As explained in \cite{atkins2014}, finding the asymptotic rate of a code with
non-linear constraints when the block length goes to infinity while maintaining
fixed constraint and variable node degree distributions is not as easy as it
is is for linear codes. For linear codes, the so-called ``design rate'' is 
a direct function of the degree distributions, and it is easy to show that
the true rate for randomly chosen graphs is unlikely to deviate much from
the design rate. In \cite{atkins2014}, we conjectured an expression for 
a rate estimate but were unable to give it a full justification. As it turns 
out, this estimate was not accurate at all. We can now make the following
precise statement:
\begin{lemma}
If a factor graph with regular constraint node degree $q$ equal to the alphabet
size is a tree, then the rate of the corresponding code tends to $\log((q-1)!)/(q-1)$
as the number of nodes tends to infinity.
\end{lemma}
{\em Proof:} Start drawing the tree from its root with one constraint node and its
associated $q$ variables. The combination of variables can take on $q!$ values.
Now add constraint nodes to any variable in the tree one at a time. With each
addition, the combination of new $q-1$ variables added can take on $(q-1)!$
values. Hence the set of values for a tree with $k$ constraint nodes is 
$q!((q-1)!)^{k-1}$ for a number of variables $q+(k-1)(q-1)$, which tends
to the expression given for $k\longrightarrow\infty$.
\qed
While this statement is correct, it is of little use for estimating the asymptotic
rate of the code. We were mislead by the convergence theorem of density
evolution, which states that, for a finite number of iterations, the horizon of a
variable node converges to a cycle-free graph as the block length grows to infinity.
While this is true for a finite number of iterations, 
it obviously does not remain true when the number of iterations grows to infinity,
which would be necessary for the rate of the overall code to converge towards the
rate of a tree. We now call this quantity the cycle-free rate $R_{cf}$.

Furthermore, the threshold calculations for variable degree $d_v=3$ codes
in \cite[Table III]{atkins2014} were erroneous, and we state the correct values 
in Table~\ref{table:thresholds} along with $1-R_{cf}$.
\begin{table}
\centering
\caption{BP thresholds for regular $(3,q)$ graphs}
\label{table:thresholds}
\begin{tabular}{|l||c|c|c|c|c|c|}
\hline
$q$ & 3 & 4 & 5 & 6 & 7 & 8 \\ \hline
Threshold $\theta$ & 
0.8836 &
0.7251 &
0.6209 &
0.5492 &
0.4965 &
0.4559 \\ \hline
$1-R_{cf}$ &
  0.6845
& 0.5692
& 0.5063
& 0.4656
& 0.4365
& 0.4143 \\
\hline
\end{tabular}
\end{table}
Beyond $q=8$, it becomes infeasible to compute thresholds
using density evolution with our limited computational resources.
We note that the thresholds are greater than $1-R_{cf}$, 
which goes to show that $R_{cf}$ is a bad estimate of the true rate
since we would otherwise be violating Shannon's converse coding theorem.

In \cite{vontobel2014}, Vontobel was able to obtain what he believes
to be an accurate estimate of the rate of a regular $(d_v,q)$ factor graph with
permutation constraints using the Bethe approximation of the partition
function of the factor graph. This approach yields
\[
R_{est} = \max\left(0, (d_v/q) \cdot \log_2(q!) - (d_v-1) \cdot \log_2(q)\right),
\]
or using Stirling's approximation $ q! \approx \sqrt{2\pi q} (q/e)^q$,
\[
R_{est} \approx \max\left(0, \log_2(q \cdot (2\pi q)^{d_v/(2q)} / e^{d_v})\right).
\]
This estimate is zero for all alphabet sizes $q$ up to and including 11. As a 
result, the asymptotic analysis of codes with local permutation constraints is
caught between a rock and a hard place: there are sub-exponentially many codewords
for all alphabet sizes up to 11, but we are unable to perform density evolution
for alphabet sizes larger than 8.

\section{Universal encoding and prefix reservation}
\label{sec:enc}

We now return to finite length codes for which the considerations of the previous
section do not apply. 
The rate of a finite length $N$ code with non-linear constraints is $R=(\log M)/N$,
where  $M$ is the number of sequences that fulfill the constraints. 
The number of valid classsical $q\times q$ SUDOKU grids for various $q$ has been counted or estimated,
as reported in \cite{math_of_sudoku}, yielding, for $q=9$, 
$M_9$=6,670,903,752,021,072,936,960 and for $q=16$, $M_{16}\approx 5.9584\times 10^{98}$,
resulting in rates of $R_9=(\log_9M_9)/81=0.28$ and $R_{16} \approx 0.32$.
The number $M_9$ was obtained through tedious counting of all valid SUDOKU grids
after excluding as many symmetries as possible. This process yields a
valid enumeration of SUDOKUs that could potentially be used for encoding purposes. 

\begin{figure}
\centering
\begin{tikzpicture}[scale=0.8]
\draw [fill=green!20] (0,3) rectangle (1.5,4) node [pos=.5] {Source};
\draw [->] (1.5,3.5) -- (3.5,3.5);
\node [below] at (2.5, 3.5) {1,0,1,1,\ldots};
\draw [fill=yellow] (3.5,3) rectangle (5.5, 4) node [pos=.5] {\shortstack{Arithmetic \\ Encoder}};
\draw [->] (5.5,3.5) -- (7.5,3.5);
\node [below] at (6.5,3.5) {3,1,4,2,\ldots};
\draw [fill=orange!80] (7.5, 3) rectangle (9.5, 4) node [pos=.5] {\shortstack{Message \\ Selector}};
\draw [->] (8.5, 3) -- (8.5, 2.5);
\draw [fill=pink] (7.4, 1.5) rectangle (9.6, 2.5) node [pos=.5] {\shortstack{Belief \\ Propagation}};
\draw [->] (8.5, 1.5) -- (8.5, 1.1);
\draw (8.5, 1.1) -- (8.4, 1) -- (8.5, .9) -- (8.6, 1) -- (8.5, 1.1);
\draw [->] (8.4,1) -- (8, 1);
\node [left] at (8, 1) {Encoding Failure};
\draw [->] (8.5, .9) -- (8.5, .5);
\node [below] at (8.5,.5) {Successful Encoding};
\draw [->] (8.6,1) -- (10,1) -- (10, 3.5) -- (9.5, 3.5);
\draw [->] (8.5,4) -- (8.5, 4.5) -- (4.5, 4.5) -- (4.5, 4);
\end{tikzpicture}
\caption{A universal encoder structure for graphs with local constraints}
\label{fig:univ-enc}
\end{figure}
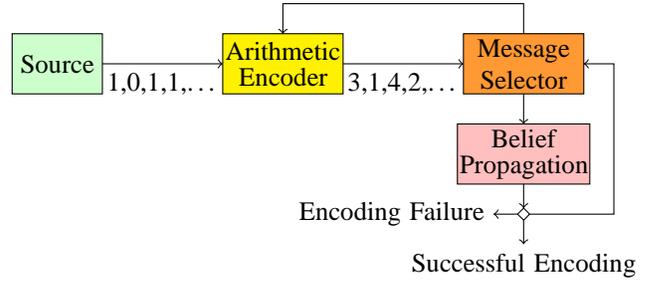
Nevertheless, we investigate a general structure illustrated in Figure~\ref{fig:univ-enc}
that can be used to encode any sequence with local constraints.
It is inspired by Richardson and Urbanke's method to encode LDPC
codes \cite{ru01_enc} using an erasure decoder. However, unlike in \cite{ru01_enc}
it is not possible to start encoding by assigning information symbols to the
systematic part of the codeword. There is no equivalent to the systematic property
for general non-linear codes, and certainly for permutation constraints there is
no part of a codeword beyond a single variable that can be assigned at will without taking
constraints into account. Thus, our approach in Figure~\ref{fig:univ-enc} starts
with an empty grid. At every step, belief propagation is used to
exclude incompatible values for the undetermined variables. Following that, the message selector
takes the first variable whose set of possible values has
cardinality $k>1$, and asks an arithmetic encoder to
convert source randomness into a uniform $k$-ary variable and uses this to determine
the value of the free variable. The reason for using an arithmetic encoder
is that the cardinality $k$ may vary
from one step to the next, while the source has a constant alphabet.

Let us for example encode a $4\times 4$ classic SUDOKU grid.
The first decoding operation on the empty grid makes no advance at all,
and we begin by requesting a choice of cardinality 4 from the arithmetic encoder
to determine the top left variable. Say the value assigned is 3, then the decoder
will automatically exclude the value 3 from all remaining variables in the first 
row, first column, and first $2\times 2$ subsquare. After this, the first available
variable is the second element of the first row whose value can be 1, 2 or 4 and 
hence we request a uniform choice between 3 possibilities from the arithmetic encoder.
Say this choice is 3, so we pick the third possible element 4 for our variable.
This continues until the grid is filled. With this method, the encoding rate may
vary from one codeword to the next and the true rate is the average of the 
individual codeword rates. 

The problem with the method proposed is that it will not always succeed. If we
had access to an optimal erasure decoder, the predictions for the possible
values of a variable would always be accurate. Belief propagation on the
other hand is a sub-optimal erasure decoder that treats each local constraint
independently. The consequence of this is that the decoder may for example
predict three possible values for a given variable, say 1, 4 and 7 in a 9-ary
alphabet, but one of these possibilities, say 1, is in fact illegal as there exists
no codeword that combines the determined values with a 1 in this position.
An optimal decoder would never propose 1 as a possible value in this context. With a
sub-optimal decoder, when a wrong prediction occurs, we are left in a precarious
situation. Unlike SUDOKU puzzle solvers who can then simply backtrack and 
follow one of the other possibilities, we cannot do this because the
decoder will never know that there was an illegal option among the 3. 
Decoding may fail because the decoder misinterprets the mapping
of information sequence to codeword, even if the transmitted codeword was
decoded correctly. Hence, the encoding process may terminate
in an encoding failure, as indicated in Figure~\ref{fig:univ-enc}.

A way around this is to reserve
a prefix of the tree of possible codewords for subsequent encoding attempts. For 
example in a $9\times 9$ SUDOKU, we may exclude the value 9 for the first variable,
effectively constraining the encoder to make a choice between 8 possibilities for
this variable. Should the encoding process result in an encoding failure, all 
source symbols consumed in this encoding operation are returned to the source, 
and encoding begins again but is preceded by the reserved prefix 9. The decoder
recognises that the reserved prefix has been used, hence knows that no source
symbol has been consumed to produce the first code symbol. This techique can be
applied recursively for example by reserving a prefix within the prefix for the
case when the encoder fails again, and so forth until the probability of an 
encoding failure becomes negligible. Furthermore, although the prefix may appear
wasted in terms of rate, it can in theory be balanced exactly against the 
probability of failure so that the resulting encoder is rate optimal. Another
way of seeing this is that the sub-optimal erasure decoder overestimates
the size of the code tree. We compensate for this by reducing
the tree by the reserved prefix. 

This prefix reservation approach is practical if the probability of encoding
failure is quite small. We will see in the next section that the probability
of an encoding failure for $9\times 9$ SUDOKU is about 1.6\%, making them 
very suited for the encoding technique we described in this section.

\section{Simulation of SUDOKU and semi-pandiagonal squares}
\label{sec:sim}

There are a number of regular graph structures for finite code design
that we have considered in our work:

{\em Latin squares:} $q^2$ variables over an alphabet of size $q$. 
  The constraints are best visualised by arranging the variables
  as a $q\times q$ square and correspond to all rows and columns.

{\em SUDOKU:} $q^2$ variables over an alphabet of size $q$, where
  $\sqrt{q}$ must be an integer. A Latin square
  with extra constraints corresponding to the
  $q$ subsquares of dimension $\sqrt{q}\times\sqrt{q}$.

{\em Pandiagonal Latin squares:} a Latin square with further
  permutation constraints corresponding to the broken right {\em and
  left} diagonals \cite{hedayat75} $(i,j+i)$ and $(i,j-i-1)$ for
  $i=0,1,\ldots,q-1$. These are also called Knut Vik designs.

{\em Semi-pandiagonal Latin squares:} a Latin square satisfying
  the broken right
  diagonal $(i,j+i)$ constraints. These have also been called Semi
  Knut Vik designs in \cite{hedayat75}.

As shown in \cite{hedayat75}, there are no semi-pandiagonal Latin squares
for $q$ even. We have counted the semi-pandiagonal Latin squares
for various $q$ and obtained $3!$ for $q=3$, $3\times 5!$ for $q=5$,
$635\times 7!$ for $q=7$, and 489,300$\times 9!$ for $q=9$.

In our simulations, we focused on SUDOKU and on semi-pandiagonal Latin squares,
mainly because for $q=9$, both these correspond to regular (9,3) factor graphs,
differing only on the last 9 of 27 constraints.
We were surprised to find that, despite these apparent topological similarities,
the properties of the code varies greatly.
The table below shows the rate and probability of encoding failure for
the two cases:
\begin{center}
\begin{tabular}{|l|c|c|}
\hline
Factor Graph & SUDOKU & Semi-pandiagonal \\ \hline

True Rate
&
\textcolor{blue}{$R=0.2824$}
&
\textcolor{red}{$R=0.1455$}
\\ \hline
Probability of &  & \\
Encoding Failure
&
\textcolor{blue}{0.016}
&
\textcolor{red}{0.9995}
\\ \hline
\end{tabular}
\end{center}
We see that semi-pandiagonal Latin squares have a lower rate and
an extremely high probability of encoding failure close to 1, making
them unsuitable for the type of encoding suggested in the previous
section. However, there is a rich literature on enumerating pandiagonal
Latin squares \cite{hedayat77, atkin83, bell07, dabbaghian13} and
it may be possible to apply some of these techniques to semi-pandiagonal
Latin square, thereby providing a method to encode them without using 
the factor graph or belief propagation on the erasure channel. 

\begin{figure}[t]
\centering
\begin{tikzpicture}[scale=.4]
\begin{axis}[%
width=6.02303149606299in,
height=4.75042322834646in,
scale only axis,
xmin=0,
xmax=0.9,
xlabel={Erasure Probability},
xmajorgrids,
ymode=log,
ymin=1e-05,
ymax=1,
yminorticks=true,
ylabel={Block Error Probability},
ymajorgrids,
yminorgrids,
legend pos = south east,
legend cell align = left
]
\addplot [
color=blue,
solid,
line width=3.0pt,
]
table[row sep=crcr]{
0.4 0.284345285714286\\
0.375 0.223100428571429\\
0.35 0.170488383673469\\
0.325 0.130298291836735\\
0.3 0.0924603040816327\\
0.275 0.0656355204081633\\
0.25 0.0452633081632653\\
0.225 0.0287716589795918\\
0.2 0.0178881657142857\\
0.175 0.010253423877551\\
0.15 0.00558563612244898\\
0.125 0.00266592797959184\\
0.1 0.00104067771428571\\
0.075 0.000334324226530612\\
0.05 5.59186897959184e-05\\
};
\addplot [
color=red,
dashed,
line width=3.0pt,
]
table[row sep=crcr]{
0.9 1.00083333333333\\
0.85 0.994995\\
0.8 0.915916\\
0.75 0.668502166666667\\
0.7 0.335335166666667\\
0.65 0.126981033333333\\
0.6 0.0501385333333333\\
0.55 0.0227092733333333\\
0.5 0.013157397\\
0.45 0.00593117001666667\\
0.4 0.00337034667\\
};
\addplot [
color=blue,
solid,
forget plot
]
table[row sep=crcr]{
0.7176 1e-05\\
0.7176 1\\
};
\addplot [
color=red,
dashed,
forget plot
]
table[row sep=crcr]{
0.8473 1e-05\\
0.8473 1\\
};
\legend{$9\times 9$ SUDOKU code,$9\times 9$ Semi-pandiagonal code,,}
\end{axis}
\end{tikzpicture}%
\caption{Block error performance of a SUDOKU and a semi-pandiagonal code. 
Each data point was averaged over 100 codewords and each codeword was subjected to a sufficient number of
erasure sequences to observe at least 100 block errors. 
}
\label{fig:sim}
\end{figure}
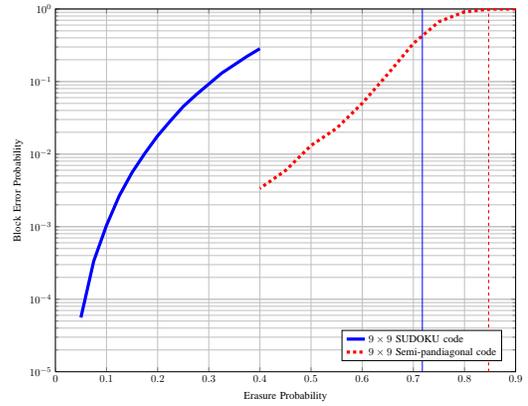

The simulated performance of both codes on the erasure channel
is shown in Figure~\ref{fig:sim}. 
The results show that the codes again have very different characteristics, with the semi-pandiagonal Latin square's error
performance flattening out while the SUDOKU exhibits a steeper waterfall
but overall further from its threshold $1-R=0.72$.

\section{Conclusion}
We have described 
non-linear codes with 
local permutation constraints inspired by SUDOKU puzzles. For the decoder,
we showed that permutation constraints require the evaluation of a 
permanent using a trellis.
We specialised this decoder to erasure channels, where the operation becomes a
trellis search. For the encoder, we described a universal
approach to encoding a code with local constraints, and discussed its
limitations when based on a sub-optimal decoder. We also gave some
insight regarding asymptotic performance, and simulation results
for select finite length codes.

Having started as a tool for teaching belief propagation,
the study of codes with non-linear constraints is having unexpected
repercussions: it has brought up some interesting technical hurdles,
taught us a few things about non-linear constraints and how different
they behave from linear constraints, and in the process also highlighted
a few properties that we take for granted in linear codes but are
in fact quite surprising. In future work, it is worth
paying more attention to structures such as pandiagonal Latin squares
for which enumerations have been devised. Another line of enquiry are
improved decoding rules that combine constraints in belief propagation.

\section*{Acknowledgment}

The authors wish to thank Gottfried Lechner for suggesting to compute permanents
with a trellis, as described in Section~\ref{sec:perm}.


\bibliographystyle{IEEEtran}
\bibliography{IEEEabrv,myrefs}
\end{document}